\documentclass[sigconf]{acmart}
\makeatletter
\renewcommand\@formatdoi[1]{\ignorespaces}
\makeatother

\usepackage{booktabs}
\usepackage{subcaption}
\usepackage{commath}
\usepackage{threeparttable}
\usepackage{caption}
\usepackage{tabularx}
\usepackage{natbib}
\usepackage{balance}

\begin{document}


\title{Mixture-of-tastes Models for Representing Users with Diverse Interests}
\author{Maciej Kula}
\email{maciej.kula@gmail.com}
\date{\today}
\settopmatter{printacmref=true, printccs=false, printfolios=true}
\setcopyright{rightsretained}
\acmConference[Learn-IR]{Workshop on Learning From User Interactions, WSDM 2018}{February 5--9, 2018}{Los Angeles, United States}



\begin{abstract}
Most existing recommendation approaches implicitly treat user tastes as unimodal, resulting in an average-of-tastes representation when multiple distinct interests are present. We show that appropriately modelling the multi-faceted nature of user tastes through a mixture-of-tastes model leads to large increases in recommendation quality. Our result holds both for deep sequence-based and traditional factorization models, and is robust to careful selection and tuning of baseline models. In sequence-based models, this improvement is achieved at a very modest cost in model complexity, making mixture-of-tastes models a straightforward improvement upon existing baselines.
\end{abstract}

\maketitle

\section{Introduction}
Latent vector-based recommender models commonly represent users with a single latent vector per user \citep{koren2009bellkor, hu2008collaborative}. This representation, while simple and efficient to compute, neglects the fact that users have diverse tastes, and that observed user interactions can be seen as manifestations of several distinct tastes or intents. These may either be stable facets of the user's preference (liking both horror and documentary movies, or both bluegrass and classical music), context-driven changes (preference for short-form TV content during the week but long-form cinematography during the weekend), or manifestations of phenomena like account sharing, where two different users, with correspondingly different tastes, share the same user account.

In all these cases, trying to capture the user's taste in a single latent vector is suboptimal. Firstly, it may lead to lack of nuance in representing the user, where a dominant taste may overpower more niche ones. Secondly, it may reduce the quality of item representations, especially when it leads to decreasing the separation in the embedding space between groups of items belonging to multiple tastes or genres. For illustration, documentaries and horror movies are distinct genres in the sense that most users prefer one or the other, but there are still users who like both. To represent that group of users, the genres' embeddings will have to be separated less cleanly than they would be if the model could express the concept of multiple tastes. In general, these problems are similar to that of fitting a unimodal distributional model to bimodal data; for Gaussians, the resulting fitted distribution will be a poor model of the data, and the majority of its density mass will coincide with neither of the true modes.

In this paper, we propose and evaluate representing users as mixtures of several distinct tastes, represented by distinct taste vectors. Each of the taste vectors is coupled with an attention vector, describing how competent it is at evaluating any given item. The user's preference is then modelled as a weighted average of all the user's tastes, with the weights given by how relevant each taste is to evaluating a given item.

We apply this model to both traditional implicit feedback factorization models, and to more recent models using recurrent neural networks. In both cases, our mixture-of-interests models outperform single-representation models on standard ranking quality metrics. In the case of deep recurrent sequence models, this improvement is achieved at a very modest cost to to the model complexity, making our model a straightforward improvement over existing methods.

\section{Related work}
The idea of moving beyond point embeddings has yielded some interesting results, particularly in the natural language modelling domain.

\citet{vilnis2014word} embed words as Gaussian distributions, rather than point embeddings. This improves the resulting representations in two ways. Firstly, the resulting representation provides a natural way of expressing uncertainty about the learned representation. Secondly, large (and non-spherical) variances aid the representation of polysemous words. In the context of recommendations, Gaussian embeddings with large variances could be used to represent users with wide-ranging tastes.

\citet{athiwaratkun2017multimodal} extend this idea by representing words as mixtures of Gaussians. This allows them to capture polysemy by modelling each word with several distinct, small-variance Gaussian distributions (rather than artificially inflating the variance of a single distribution). This allows much clearer separation of distinct meanings.

In the recommender system literature, the idea of using multiple embeddings to represent a user's multiple interests is presented in \citet{weston2013nonlinear}. In this approach, the recommendation score of an item for a given user is given by the \emph{maximum} of the dot products of the item embedding and each of the user's embedding vectors. This obviates the need for explicit modelling of mixture probabilities, which reduces the number of model parameters and makes evaluation more efficient. However, the model is potentially disadvantaged by its inability to model strong \emph{dis}taste for a given class of items.

The approach closest to the one presented here, but focusing solely on the sequence-based setting and using pre-trained item embeddings, is presented in \citet{wang2017attention}, and was published contemporaneously with the writing of this paper. The authors use recurrent neural networks with attention to model user sequences, and either feedforward or recurrent decoders to generate taste mixture components. Their experiments confirm that the addition of taste mixtures improves recommendation quality. We suggest that the two papers should be seen as mutually reinforcing and complementary, and that a combination of both approaches (recurrent decoders and end-to-end embedding training) would be beneficial.

\section{Model}
We propose three variants of the mixture-of-tastes model, covering both recurrent and traditional factorization models.

Within the classical matrix factorization framework, we experiment with two approaches to modelling taste mixtures. The first is the Embedding-Mixture Factorization (\textbf{EM-F}) model. In this model, a user is represented by two sets of $m$ $k$-dimensional latent vectors: one where the vectors correspond to their tastes, and one that describes how competent each taste is at describing a given item. When evaluating an item, the dot products of the second set of vectors with the item latent vector give the relative weights of each tastes from the first set that should be applied when evaluating the user's preference. Compared to a traditional factorization model, the EM-F model requires $2m$ as many parameters, giving it more expressive power at the expense of using more parameters.

The second factorization model we test is the Projection-Mixture Factorization (\textbf{PM-F}) model. It extends the basic matrix factorization model by using linear projection layers in conjunction with latent vectors to model multiple tastes: instead of embedding $m$ taste vectors and $m$ mixture vectors for each user directly, we use $2m$ $k \times k$ projection matrices to project a single $k$-dimensional user item vector into $m$ taste and $m$ mixture vectors. These matrices are shared between all users, making the model substantially more parsimonious.

Our sequence-based model is the Mixture LSTM (\textbf{M-LSTM}) model. It builds on prior work \citep{wu2017recurrent, hidasi2015session} using recurrent neural networks to model the sequence of user interactions. In our model, we use a long-short term memory network (LSTM, \cite{hochreiter1997long}) to transform the sequence of user interactions into a latent representation; we then project it into the item embedding space via a number of linear projection layers to obtain the mixture-of-tastes representation. By using a recurrent architecture, we capture information from the identity of the items the user interacted with as well as the order in which those interactions occurred.

Formally, let $U_i$ be a $m \times k$ matrix representing the $m$ tastes of user $i$, and $A_i$ be a $m \times k$ matrix representing the affinities of each taste from $U_i$ for representing particular items. The recommendation score for item $j$, whose representation is given by a $k \times 1$-dimensional embedding vector $e_j$, is then given by
\begin{equation}
  r_{ij} = \sigma\left(A_ie_j\right) \cdot U_ie_j + b_j,
\end{equation}
where $b_j$ is the per-item bias term, $\sigma$ the softmax function, $\sigma\left(A_ie_j\right)$ gives the mixture probabilities, $U_ie_j$ the recommendation scores given by each mixture component, and $\cdot$ denotes the vector dot product. We assume identity variance matrices for all mixture components.

How $U_i$ and $A_i$ are obtained differs between the three evaluated models. In the M-LSTM model, $U_{it}$ and $A_{it}$ (now indexed by $t$, their position in the sequence of interactions) are linear functions of the hidden state $z_{it}$ of an LSTM layer trained on the user's previous interactions:
\begin{equation}
  \label{eq:lstm}
  z_{it} = \mathrm{LSTM}\left(e_{i1}, e_{i2}, \ldots, e_{it}\right).
\end{equation}
Given the $1 \times k$ dimensional hidden state $z_{it}$, the $m$-th row of $U_{it}$ and $A_{it}$ are given by
\begin{equation}
\begin{aligned}
  u^m_{it} &= z_{it}W^U_m + B^U_m\\
  a^m_{it} &= z_{it}W^A_m + B^A_m,\\
\end{aligned}
\end{equation}
where $W^U_m$, $W^A_M$ are the learned projection matrices, and $B^U_m$ and $B^A_m$ contain bias terms. Both the projection matrices and the biases are common across all users, representing a modest increase in the total number of model parameters. Note that the LSTM network only needs to be run once to obtain the full user representation.

The PM-F model is similar: an embedding $z_i$ is estimated for each user, and the $U_i$ and $A_i$ matrices are obtained via linear projections:
\begin{equation}
\begin{aligned}
  u^m_{i} &= z_{i}W^U_m + B^U_m\\
  a^m_{i} &= z_{i}W^A_m + B^A_m.\\
\end{aligned}
\end{equation}
This keeps the number of model parameters small at a potential cost to model expressiveness. In contrast, the EM-F model embeds $U_i$ and $A_i$ directly. This substantially increases the number of model parameters, but may lead to better accuracy.

In all models, the item representations are given by the latent vectors $e$. The input and output item embeddings are tied, and they have the same dimensionality as the user representations.

\section{Experiments}
We test our models on a number of publicly available datasets with varying degrees of sparsity as well as diversity of tastes. In all tests, we treat our models as ranking models, and evaluate them on the quality of the ranked recommendation list they generate, rather than the rating predictions they produce (unlike \cite{wu2017recurrent}).


\subsection{Datasets}
We use the following datasets in our experiments (their summary statistics are listed in Table \ref{tab:datasets}):

\begin{enumerate}
\item Movielens 10M: dataset of 10 million movie ratings across 10,000 movies and 72,000 users \citep{harper2016movielens}.
\item Goodbooks-10K: dataset of 6 million ratings across 53,000 users and 10,000 most popular books from the Goodreads online book recommendation and sharing service \citep{goodbooks2017}.
\item Amazon: dataset of ratings and reviews gathered from the Amazon online shopping service, spanning books, music, and videos \citep{leskovec2007dynamics}. After pruning users and items with fewer than 10 ratings, the dataset contains approximately 4 million ratings from 100,000 users over 114,000 items.
\end{enumerate}
The two key differences between the datasets are their sparsity and the degree to which they are popularity-biased. The Movielens 10M and Goodbooks datasets are relatively dense, while the Amazon dataset is much sparser. Popularity seems to play a greater role in the Movielens dataset than in the other datasets: the ratio between the number of interactions that accrue to the 95th percentile and the 50th percentile of most popular items is highest in Movielens 10M.

We conjecture that taste diversity plays a lesser role in highly popularity-biased datasets (that is, where a large share of all interactions go to very few items). Intuitively, a single dominant taste can model such observations quite well; additional tastes only come into play in the long tail of the distribution. Conversely, a more even distribution of interactions between items allows the possibility that multiple tastes play a larger role. If this is true, we would expect the gains from our mixture models to be larger in the Amazon (where tastes can span multiple unrelated item categories) and Goodbooks (where the popularity distribution is relatively even by construction) datasets than in the Movielens dataset.

Throughout our experiments, we treat all datasets as implicit feedback datasets, where the existence of an edge between a user and an item expresses implicit preference, and the lack of an edge implicit lack of preference.

\begin{table}
  \caption{Dataset statistics. 95th/50th denotes the ratio of popularity of the 95th and 50th percentile of item popularity.}
  \label{tab:datasets}
\begin{tabularx}{\columnwidth}{lllrr}
\toprule
 Dataset       & Users   & Items   &   Density &   95th/50th \\
\midrule
 Movielens & 69,879  & 10,678  &    0.0134 &      7.42 \\
 Amazon        & 100,085 & 113,997 &    0.0003 &      5.67 \\
 Goodbooks     & 53,425  & 10,001  &    0.0112 &      1.41 \\
\bottomrule
\end{tabularx}
\end{table}

\subsection{Baselines}
Our baselines are exact equivalents of the mixture models, differing only in the fact that they represent the user with a single $k$-dimensional vector. For sequence-based models, we use an LSTM architecture and represent the user directly with the last hidden state of the network ($z_{it}$ from equation \ref{eq:lstm}). For factorization models, we use a standard latent embedding vector, corresponding directly to $z_i$ from the projection mixture model.

This makes our baselines particularly suitable for evaluating mixture-of-tastes representations: adding multiple tastes is the sole architectural difference between the models, and so differences in recommendation performance can be directly attributed to the models' greater ability to model diverse tastes.

\subsection{Experimental setup}
For factorization models, we split the interaction datasets randomly into train, validation, and test sets. We use 80\% of interactions for training, and 10\% each for validation and testing. We make no effort to ensure that all items and users in the validation and test sets have a minimum number of training interactions. Our results therefore represent partial cold-start conditions.

For sequence-based models, we order all interactions chronologically, and split the dataset by randomly assigning users into train, validation, and test sets. This means that the train, test, and validation sets are disjoint along the user dimension. For each dataset, we define a maximum interaction sequence length. This is set to 100 for the Goodbooks and Movielens datasets, and 50 for the Amazon dataset, as the interaction sequences in the Amazon dataset are generally shorter. Sequences shorter than the maximum sequence length are padded with zeros. The models are trained by trying to predict the next item that the user interacts with on the basis of all their prior interactions.

We use mean reciprocal rank (MRR) as our measure of model quality. In factorization models, we use the user representations obtained from the training set to construct rankings over items in the test set. In sequence models, we use the last element of the test interaction sequence as the prediction target; the remaining elements are used to compute the user representation.

We believe our sequence model experimental setting to be a relatively good reflection of the conditions in which industry recommender systems are trained and evaluated: model retraining is often done daily, with data up to the day of training used for model estimation, and subsequent interactions used for evaluation. Insofar as this is true, our results should generalize well to real-world applications.

\subsection{Loss functions}
We experiment with two loss functions:
\begin{itemize}
\item Bayesian personalised ranking (BPR, \citet{rendle2009bpr}), and
\item adaptive sampling maximum margin loss, following \citet{weston2011wsabie}.
\end{itemize}
For both loss functions, for any known positive user-item interaction pair $(i, j)$, we uniformly sample an implicit negative item $k$. For BPR, the loss for any such triplet is given by
\begin{equation}
1 - \sigma\left(r_{ij} - r_{ik}\right),
\end{equation}
where $\sigma$ denotes the sigmoid function.
The adaptive sampling loss is given by
\begin{equation}
\abs{1 - r_{ij} + r_{ik}}_{+}.
\end{equation}
For any $(i, j)$ pair, if the sampled negative item $k$ results in a zero loss (that is, the desired pairwise ordering is not violated), a new negative item is sampled, up to a maximum number of attempts. This leads the model to perform more gradient updates in areas where its ranking performance is poorest.

Across all of our experiments, the adaptive maximum margin loss consistently outperforms the BPR loss on both baseline and mixture models. We therefore only report results for the adaptive loss.

\subsection{Hyperparameter search}

We perform extensive hyperparameter optimization across both our proposed models and all baselines. Our goal is two-fold. Firstly, we want to mitigate researcher bias, where more care and attention is devoted to the researcher's proposed model, thus unfairly disadvantaging baseline algorithms. We believe this to be a common phenomenon; its extent is illustrated, in a related domain, by \cite{melis2017state}, who find that standard LSTM architectures when properly tuned outperform more recent algorithms in natural laguage modelling tasks. Secondly, we wish to understand the extent to which the mixture-of-interests models are \emph{fragile}, in the sense of being highly sensitive to hyperparameter choices. Such fragile algorithms are potentially of lesser utility in industry applications, where the engineering cost of tuning and maintaining them may outweigh the accuracy benefits they bring.

We use random search to tune the algorithms used in our experiments. We optimize batch size, number of training epochs, the learning rate, L2 regularization weight, the loss function, and (where appropriate) the number of taste mixture components.

\subsection{Implementation}
Our models are implemented using the PyTorch deep learning framework \citep{paszke2017pytorch} and trained using the nVidia K40 GPUs. All of the models are trained using the ADAM \citep{kingma2014adam} per-parameter learning rate schedule. We make the model and experiment code (as well as the full results) available on Github\footnote{\url{https://github.com/maciejkula/mixture}}.

\section{Results}
Our main results are summarized in Table \ref{tb:results}. In both sequence-based and factorization tasks, variants of our model achieve consistently higher performance than baseline models, with improvements ranging from 10\% to 69\%. The results are robust to optimizing the hyperparameters of our baselines, giving us ample confidence that the results can be replicated in production recommender systems.
\begin{table}
\caption{Mean reciprocal rank (MRR) across all users/sequences in the test set. Note that due to differences in experimental protocol, results between sequence-based and factorization models are not directly comparable.}
\label{tb:results}
\begin{subtable}{\columnwidth}
\caption{Sequence models}
\begin{tabularx}{\columnwidth}{lrrr}
\toprule
 Model        &   Movielens &   Amazon &   Goodbooks \\
\midrule
 LSTM         &          0.0908 &   0.1502 &      0.1158 \\
 Mixture-LSTM &          \textbf{0.1001} &   \textbf{0.1889} &      \textbf{0.1358} \\
\bottomrule
\end{tabularx}
\end{subtable}
\hspace{\fill}
\begin{subtable}{\columnwidth}
\caption{Factorization models}
\begin{tabularx}{\columnwidth}{lrrr}
\toprule
 Model              &   Movielens &   Amazon &   Goodbooks \\
\midrule
 Factorization           &          0.1053 &   0.1001 &      0.0738 \\
 Projection Mixture &          \textbf{0.1097} &   0.0698 &      0.0712 \\
 Embedding Mixture  &          0.1026 &   \textbf{0.1696} &      \textbf{0.0853} \\
\bottomrule
\end{tabularx}
\end{subtable}

\end{table}

\subsection{Ranking quality}
  \begin{figure*}[h!]
    \centering
    \captionsetup{width=.8\linewidth}
    \caption{Maximum test MRR vs number of hyperparameter search iterations. Sequence-based models in the top row; factorization-based models in the bottom row.}
    \label{fig:hyper}
    \includegraphics[width=0.8\textwidth]{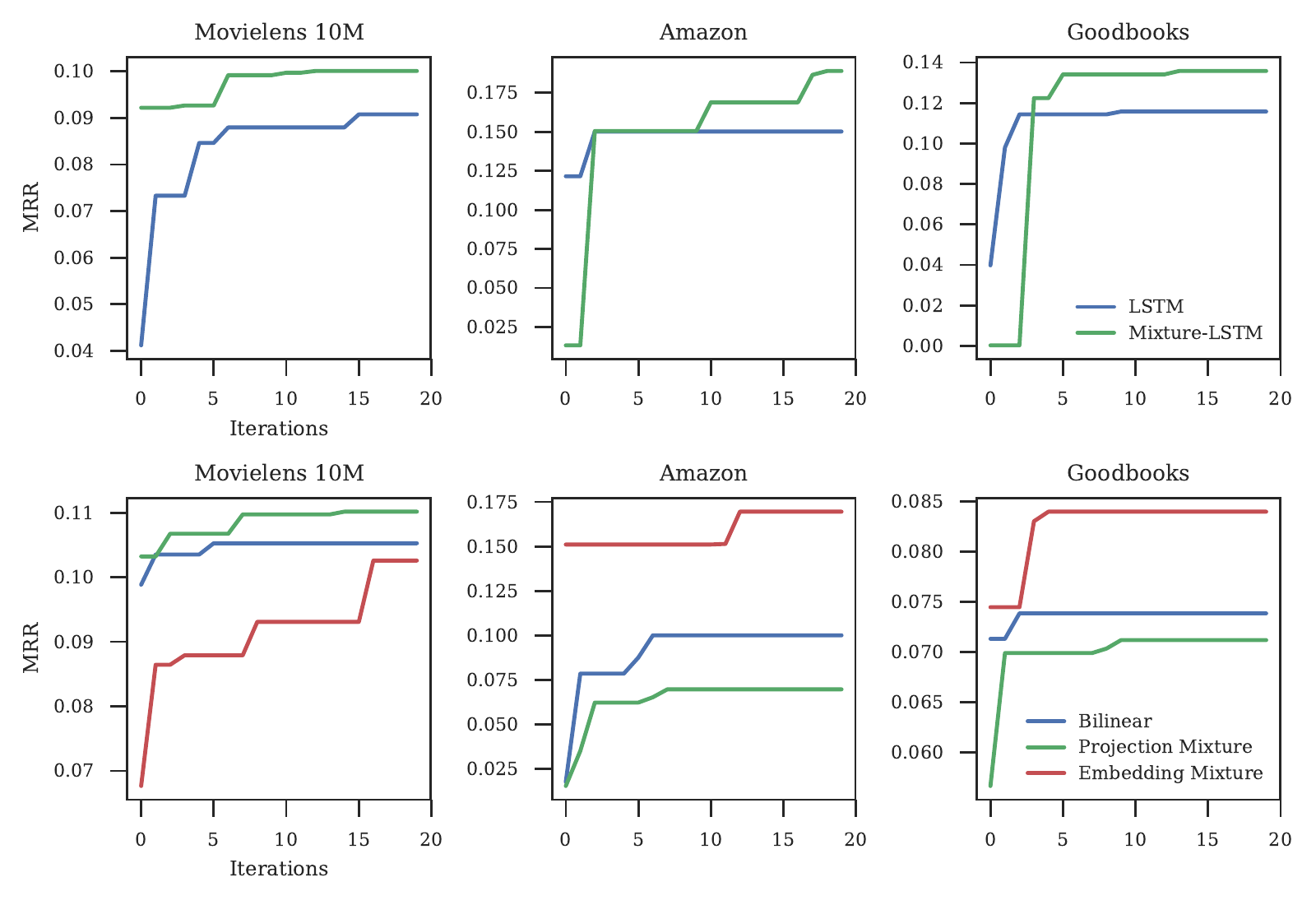}
  \end{figure*}
In sequence-based models, our M-LSTM model outperforms the baseline LSTM model on all datasets. The performance gains are particularly large on the Amazon (26\%) dataset and Goodbooks (17\%) datasets, and smaller, but still meaningful, on the Movielens 10M dataset (10\%). This is consistent with our conjectures on the nature of the datasets. The Amazon dataset, spanning the entire catalog of Amazon products, benefits most from being able to model taste diversity. This is not surprising, given that a single Amazon user can interact with products from many disparate domains.

Among factorization models, the Embedding Mixture model is the best performing model in two datasets. It beats the baseline model by a very substantial margin on the Amazon dataset (69\%), and by a smaller margin of 12\% on the Goodbooks dataset. There is relatively little to distinguish the models on the Movielens dataset, where all three models perform equally well. Somewhat surprisingly, the Projection Mixture model fails to outperform the baseline on two datasets, and substantially \emph{underperforms} it on the Amazon dataset. Given that the model is strictly more expressive than the baseline model, we attribute this failure to difficulties in effectively fitting the model.

\subsection{Hyperparameter search}
Our results are robust to hyperparameter optimization. Figure~\ref{fig:hyper} plots the maximum test MRR achieved by each algorithm as a function of the number of elapsed hyperparameter search iterations. Both baseline and mixture models benefit from hyperparameter tuning. All algorithms converge to their optimum performance relatively quickly, suggesting a degree of robustness to hyperparameter choices. Mixture-LSTM and Embedding Mixture models quickly outperform their baseline counterparts, and maintain a stable performance lead thereafter (with the exception of the factorization Movielens experiments). This lends support to our belief that the mixture models' superior accuracy reflects their greater capacity to model the recommendation problem well, rather than being an artifact of the experimental procedure or researcher bias.

\subsection{Number of taste components}
\begin{table}
\caption{Effect of number of mixture components}
\label{tab:nummixtures}
\begin{subtable}{\columnwidth}
\caption{Sequence models}
\begin{tabularx}{\columnwidth}{rrrr}
\toprule
   Components &   Movielens 10M &   Amazon &   Goodbooks \\
\midrule
            2 &          0.0882 &   0.1538 &      0.1262 \\
            4 &          0.0997 &   0.1689 &      0.1304 \\
            6 &          0.0922 &   \textbf{0.1889} &      \textbf{0.1358} \\
            8 &          \textbf{0.1001} &   0.1865 &      0.1327 \\
\bottomrule
\end{tabularx}
\end{subtable}
\hspace{\fill}
\begin{subtable}{\columnwidth}
\caption{Factorization models}
\begin{tabularx}{\columnwidth}{rrrr}
\toprule
   Components &   Movielens 10M &   Amazon &   Goodbooks \\
\midrule
            2 &          0.0931 &   0.1343 &      0.0797 \\
            4 &          \textbf{0.1026} &   0.1583 &      0.0840 \\
            6 &          0.0864 &   0.1612 &      0.0811 \\
            8 &          0.0879 &   \textbf{0.1696} &      \textbf{0.0853} \\
\bottomrule
\end{tabularx}
\end{subtable}\end{table}
We summarize the effect of increasing the number of taste mixture components in Table~\ref{tab:nummixtures}. The optimum number of components is dataset and algorithm dependent but, by and large, there is a dose-response relationship between the number of mixtures and recommendation quality: being able to represent more distinct user tastes yields better results.

\section{Conclusion}
We show that mixture-of-tastes representations are clear improvements over their baseline modes, especially in sequence-based settings where the accuracy gains come at a very modest cost in model complexity. We have taken care to test the mixture models against strong baselines, and have confidence that our approach can translate to accuracy gains in production recommender systems. We believe that modelling the diversity of user tastes in this manner can become a standard component of recommendation algorithms.

\bibliography{bibliography}
\bibliographystyle{ACM-Reference-Format}

\end{document}